\documentclass{aa}
\usepackage[colorlinks=true,     linkcolor=blue, citecolor=blue, filecolor=blue, urlcolor=blue]{hyperref}

\usepackage{color}
\usepackage{subfigure}
\usepackage{longtable}
\usepackage{multicol}
\usepackage{gensymb}

\usepackage{txfonts}
\begin{document}

   \title{Candidate ram-pressure stripped galaxies in six low-redshift clusters revealed from ultraviolet imaging}

   \authorrunning{Koshy George\inst{1}\fnmsep\thanks{koshyastro@gmail.com}}

\author{Koshy George\inst{1}\fnmsep\thanks{koshyastro@gmail.com}, B. M. Poggianti\inst{2},  A. Omizzolo\inst{3,2}, B. Vulcani\inst{2}, P. C{\^o}t{\'e}\inst{4}, J. Postma\inst{5}, Rory Smith\inst{6}, Yara L. Jaffe\inst{7}, M. Gullieuszik\inst{2}, A. Moretti\inst{2}, A. Subramaniam\inst{8},  P. Sreekumar\inst{9},  S. K. Ghosh\inst{10}, S.N. Tandon\inst{11}, J. B. Hutchings\inst{4}}

\institute{Faculty of Physics, Ludwig-Maximilians-Universit{\"a}t, Scheinerstr. 1, 81679, Munich, Germany
                                                          \and
INAF-Astronomical Observatory of Padova,  vicolo dell'Osservatorio 5 35122 Padova, Italy
\and
Vatican Observatory, Vatican City, Vatican State
\and
National Research Council of Canada, Herzberg Astronomy and Astrophysics Research Centre, Victoria, Canada
\and
University of Calgary, Calgary, Alberta, Canada
\and
Departamento de Física, Universidad Técnica Federico Santa María, Vicuña Mackenna 3939, San Joaquín, Santiago de Chile
\and
Departamento de F\'isica, Universidad T\'ecnica Federico Santa Mar\'ia, Avenida Espa\~na 1680, Valpara\'iso, Chile
\and
Indian Institute of Astrophysics, Koramangala II Block, Bangalore, India
\and
Manipal Centre for Natural Sciences, Centre of Excellence, Manipal Academy of Higher Education, Manipal 576104, Karnataka, India
\and
Tata Institute of Fundamental Research, Mumbai, India
\and
Inter-University Center for Astronomy and Astrophysics, Pune, India
}



  \abstract
  { 
  
The assembly of galaxy clusters is understood to be a hierarchical process with a continuous accretion of galaxies over time, which increases the cluster size and mass. Late-type galaxies that fall into clusters can undergo ram-pressure stripping, forming extended gas tails within which star formation can happen. The number, location, and tail orientations of such galaxies provide clues about the galaxy infall process, the assembly of the cluster over time, and the consequences of infall for galaxy evolution. Here, we utilise the $\sim$ 0.5-degree diameter circular field of the Ultraviolet Imaging Telescope to image six galaxy clusters at z$<$0.06 that are known to contain ‘jellyfish’ galaxies. We searched for stripping candidates in the ultraviolet images of these clusters, which revealed 54 candidates showing signs of unilateral extra-planar emission, due to ram-pressure stripping. Seven candidates had already been identified as likely stripping based on optical B-band imaging. We identified 47 new candidates through UV imaging. Spectroscopic redshift information is available for 39 of these candidate galaxies, of which 19 are associated with six clusters. The galaxies with spectroscopic redshifts that are not part of the clusters appear to be within structures at different redshifts identified as additional peaks in the redshift distribution of galaxies, indicating that they might be ram-pressure stripped or disturbed galaxies in other structures along the line of sight. We examine the orbital history of these galaxies based on their location in the position-velocity phase-space diagram and explore a possible connection to the orientation of the tail direction among cluster member candidates. There are limitations due to different integration times and imaging different regions with respect to the cluster centre. The tails of confirmed cluster member galaxies are found to be oriented away from the cluster centre. 

  }



\maketitle
%

\section{Introduction}
According to the hierarchical paradigm of structure formation, the massive halos in the Universe form from the continuous accretion and mergers of sub-halos. Galaxy clusters, which are the most massive structures (halo mass $\sim$ 10$^{14}$--10$^{15}$ M$_{\odot}$) found in the local Universe, have thus been assembled through the continuous accretion of infalling galaxy groups and individual galaxies \citep{Lacey_1993,Lacey_1994}. These infalling galaxies can be of any morphology, gas content, and star formation state. However, the cores of rich clusters are usually dominated by early-type (elliptical or lenticular) galaxies that have little gas and no current star formation --- at odds with the more variegated properties of accreting galaxies \citep{Dressler_1980}. This suggests there should be a cluster-specific process that converts infalling, star-forming, late-type (spiral) galaxies into non-star-forming galaxies. Possible processes include ram-pressure stripping, starvation, and tidal interactions, all of which can deplete gas and subsequently quench star formation. The most prominent of these is perhaps ram-pressure stripping, a hydrodynamic process through which the hot intra-cluster medium (T $\sim$ 10$^8$ K) strips the relatively cold gas (T $\sim$ 10$^2$ K) from galaxy discs \citep{Gunn_1972}. The efficiency of such stripping varies with both the differential three-dimensional velocity of the infalling galaxy through the intra-cluster medium with respect to the cluster and the local intra-cluster medium density, which generally decreases from the centre to the outskirts (a detailed analytic treatment for this phenomena can be found in \citealt{Jaffe_2018}). The stripped gas can condense to form new stars, which are then visible as tentacles of material trailing from the disc of the galaxy, giving a ‘jellyfish’ appearance in ultraviolet (UV) and optical light \citep{Owen_2006,Cortese_2007,Smith_2010,Owers_2012,Abramson_2014,Fumagalli_2014,Ebeling_2014,Rawle_2014,Kenney_2015,Poggianti_2016,Bellhouse_2017,Gullieuszik_2017,George_2018,Boselli_2018,Cramer_2019,Boselli_2021,Giunchi_2023a,Gullieuszik_2023,Waldron_2023}. Cosmological simulations demonstrate the presence of a large fraction of jellyfish galaxies in galaxy groups and clusters over a range of distances from the centre at different redshifts \citep{Yun_2019,Zinger_2024}. \\

Searches for such ‘unilateral’ features in optical images of galaxies in rich clusters have uncovered ram-pressure stripping candidates. Follow-up studies of such objects provide insights into the processes of gas stripping, star formation in the stripped tails, and galaxy cluster assembly \citep{Poggianti_2016, McPartland_2016,Ebeling_2019, Roman_2019,Durret_2021,Roberts_2022a,Vulcani_2022}. The identification of ram-pressure stripped galaxies is often performed either by visual inspection of imaging data from ground-based telescopes \citep{Poggianti_2016,Vulcani_2022} or by measuring the quantitative morphological parameters \citep{McPartland_2016,Krabbe_2024,Gondhalekar_2024}. Based on a visual search, from Galaxy Evolution Explorer (GALEX) imaging, of morphological asymmetries in their UV emission, \citet{Smith_2010} found evidence for ram-pressure stripping in 13 galaxies in the Coma cluster. From an inspection of B-band wide-field imaging of 71 galaxy clusters at $z$=0.04-0.07, \citet[hereafter P16]{Poggianti_2016} identified 344 candidates for ram-pressure stripped galaxies; of these, 64 were subsequently observed with the Multi Unit Spectroscopic Explorer (MUSE) as part of the GAs Stripping Phenomena in galaxies (GASP) survey and confirmed to be ram-pressure stripped in at least 86\% of the cases (Poggianti et al. in prep.). A sample of 41 stripping candidates in the Coma cluster is  presented by \citet{Roberts_2020} based on a visual inspection of optical images. The features arising from star formation in stripped tails are generally found to be clumpy with very low surface brightness, and thus require high-contrast, high-sensitivity imaging for proper detection.  The visual inspection is usually done with imaging data from blue optical bands that are sensitive to recent star formation. Imaging at UV wavelengths is even more strongly dominated by radiation from stars on the main sequence — usually with ages $<$200-300 Myrs — and hence is a direct probe of  ongoing or recent star formation. Indeed, the intrinsic emission of regions with strong, ongoing star formation can be significantly higher in the UV than in the optical. Ongoing star formation within the stripped tails, as well as the intense star formation in the disc itself, can thus make these galaxies among the brightest cluster objects when observed at UV wavelengths \citep{George_2018}

Tails observed at other wavelengths can be also used to identify galaxies that are probably undergoing ram-pressure stripping. Atomic, molecular gas content, and radio continuum studies of galaxies in nearby clusters have discovered galaxies undergoing ram-pressure stripping \citep{Kenney_1989,Kenney_2004,Chung_2007,Cortese_2010,Wang_2021,Roberts_2021a,Roberts_2022b,Hess_2022,Serra_2023,Ignesti_2023}. A recent search using 144 MHz radio continuum observations from the Low Frequency Array (LOFAR) identified 95 candidate galaxies in 29 galaxy clusters \citep{Roberts_2021a}. These objects, which are mostly found in the cluster cores,  might be the sign of galaxies experiencing their first cluster crossing following infall. The estimated size of the stripped tail and the emission timescale can change with the wavelength region (e.g. radio, optical, or UV) used to analyse the stripped galaxies. This is largely due to the different physical mechanisms (i.e. thermal or non-thermal) responsible for the flux emitted from the stripped tails (which are valuable for identifying candidates). In short, the use of  multi-wavelength imaging data  benefits the search for unilateral emission, while the relative positions and orientations of stripped tails improves our understanding of infall and the gradual buildup of galaxy clusters in the Universe. 

GASP is an integral-field-unit (IFU) spectroscopic survey of jellyfish galaxies that uses MUSE on the VLT  to study gas removal processes in galaxies \citep{Poggianti_2017}. These observations target candidate jellyfish galaxies in clusters selected from the sample of P16. We are conducting UV follow-up observations to understand the nature of ongoing star formation within  the discs and tails of these galaxies. Detailed UV imaging studies of GASP jellyfish galaxies show the frequent presence of ongoing star formation within the stripped tails \citep{George_2018,Poggianti_2019,George_2023}.  The GASP jellyfish galaxies are mostly found with projected separations very close to the centres of their host galaxy clusters, which cover the redshift from $z \sim$ 0.04 to 0.07 \citep{Jaffe_2018,Gullieuszik_2020}. With a wide,  $\sim$ 0.5$\degree$ field diameter, Ultraviolet Imaging Telescope (UVIT) imaging data  (see below) can almost cover the cluster virial radii of these clusters. This provides us with a unique opportunity to identify galaxies that are being stripped of their gas content, within which star formation may be happening. These galaxies typically have asymmetric, unilateral emission features that are readily detectable in UV images.

The main goal of this paper is to present the detection of ram-pressure stripping candidate galaxies from new UV imaging of six galaxy cluster fields from GASP. We discuss the overall sample here with details on the morphological features, tail direction, and location of the galaxies within the cluster, and check for any correlation between the detection of optical and UV features in the stripped galaxies. The tail direction and the location of galaxies inside the cluster can help in understanding the timescales and effects of ram-pressure stripping in cluster environments \citep{Rory_2022}. We discuss the observations and data analysis in Section 2, and examine the position of candidate galaxies in the phase-space diagram, as well as the distribution of tail directions, in Sections 3 and 4. We summarise our key findings in Section 5. Throughout this paper, we adopt a concordance $\Lambda$ CDM cosmology with $H_{0} = 70\,\mathrm{km\,s^{-1}\,Mpc^{-1}}$, $\Omega_{\rm{M}} = 0.3$, and $\Omega_{\Lambda} = 0.7$.

\section{Observations, data, and analysis}

\begin{table*}
\caption{\label{T1} Log of UVIT imaging observations for six galaxy clusters.}
\centering
\label{galaxy details}
\tabcolsep=0.1cm
\begin{tabular}{lccccccc} 
\hline 
\hline 
Galaxy cluster & RA$_{J2000}$ & Dec$_{J2000}$ & Channel & Filter & $\lambda_{\rm mean}$  &  $\delta\lambda$  & Integration Time \\
 & (h:m:s) & ($^\circ$:$\arcmin$:$\arcsec$) & & & ({\AA})  &  ({\AA})  & (sec)  \\
 (1) & (2) & (3) & (4) & (5) & (6) & (7) & (8) \\
\hline
Abell 1668 & 13:04:06 & $+$19:17:13 & FUV  & F154W      &  1541  &   380     &  17989\\
Abell 1991   & 14:53:51 & $+$18:39:05   &  FUV  & F148W       &  1481  &   500     &  12900\\
&  &   & NUV  & N242W     &  2418  &   785     &  12516 \\
Abell 2626 & 23:36:25 & $+$21:09:03   & FUV  & F148W       &  1481  &   500     &  11539\\
&  &   & NUV  & N242W     &  2418  &   785     &  10107 \\
Abell 3376 & 06:00:48 & $-$39:55:07  & FUV  & F148W       &  1481  &   500     & 31619 \\
Abell 4059 & 23:57:01 & $-$34:40:50  &  FUV  & F148W       &  1481  &   500  & 37504 \\
Abell 85  & 00:41:50 & $-$09:19:37     & FUV  & F148W       &  1481  &   500     &  15429\\
&  &   & NUV  & N242W     &  2418  &   785     &  18326 \\
\hline
\end{tabular}
\tablefoot{\scriptsize  Key to columns: (1) Cluster name; (2-3) UVIT pointing centre coordinates;  (4) UVIT channels used; (5) filter for imaging observations; (6) mean wavelength for filter; (7) filter bandwidth; and (8) total integration time. NUV and FUV imaging data exist for three clusters; for three others, only FUV imaging is available. Note that the six clusters were imaged in NUV and FUV to varying depths.
The UVIT NUV and FUV filters used in this study are broadly similar to the NUV and FUV filters on GALEX (see \citealt{Tandon_2017}).}
\end{table*}

UV imaging observations for cluster fields centred on each of six GASP jellyfish galaxies (JO201, JW100, JO60, JW39, JO194, and JW108) were performed using UVIT onboard the AstroSat mission. Based on the presence of stripping features identified in optical B-band images, a jellyfish class ranking was assigned  by P16 to these galaxies, with JClass=5 being the most secure and JClass=1 the least.  P16 presented a catalog of 344 jellyfish galaxy candidates based on visual inspection of B and V band imaging data of 71 galaxy clusters from OmegaWINGS and WINGS surveys \citep{Fasano_2006, Gullieuszik_2015}. The galaxies JO201, JW100, JO60, JW39, and JO194 belong to JClass=5, while JW108 was assigned a classification of JClass=2. Thus, except for JW108, all of the targets feature spectacular stripped tails. JW108 appears to be in a more advanced stripping stage, with a very pronounced truncation of star formation within the disc of the galaxy.

\begin{figure}
\centerline{\includegraphics[width=0.53\textwidth]{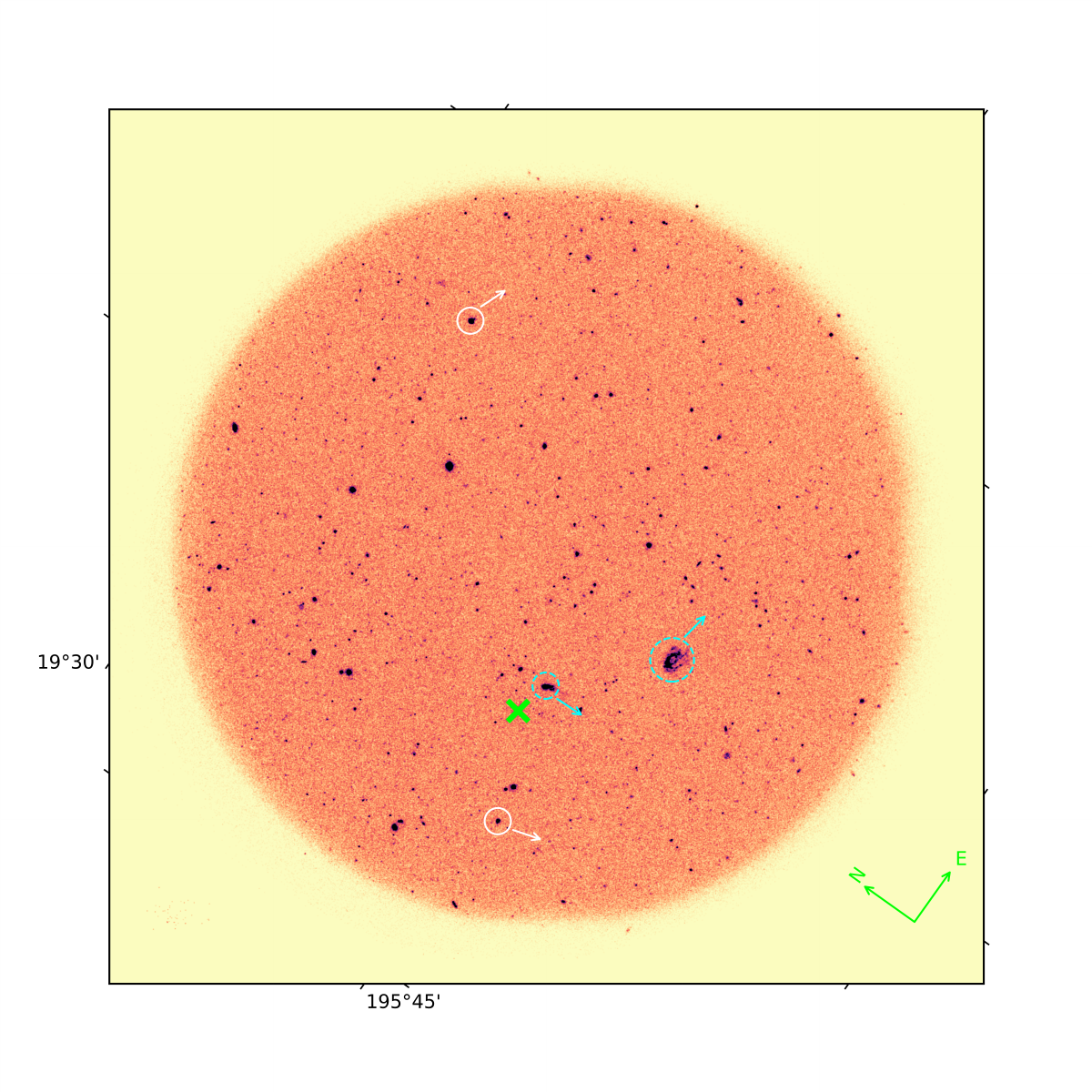}}
\caption{FUV image of Abell 1668 with the  jellyfish galaxy candidates marked as white circles and tail directions shown with arrows.  Galaxies with spectroscopic redshift information are indicated with dashed circles. Cluster members are shown with cyan circles; the GASP jellyfish galaxy is plotted with a larger (cyan) circle. The centre of the cluster is marked with a green cross. 
}\label{figure:Abell 1668}
\end{figure}

The UVIT image for one cluster field is shown here in Fig. \ref{figure:Abell 1668} and the rest in Zenodo. As was noted above, each of these clusters were imaged in optical B and V filters as part of the WINGS and OmegaWINGS surveys. We therefore include these optical images in the analysis below.

The UVIT instrument consists of twin 38-cm-diameter telescopes --- a far-ultraviolet (FUV, 130-180nm) telescope and a near-ultraviolet (NUV, 200-300nm) / VIS (320-550nm) telescope --- which operate with a dichroic beam splitter. The telescopes generate circular images over a field $\sim$ 28-arcmin in diameter, simultaneously, in all three channels \citep{Kumar_2012}.   The NUV and FUV images were corrected for distortion \citep{Girish_2017}, flat-field illumination,
and satellite drift using the latest version of the {\tt CCDLAB} \citep{Postma_2017} software package. We note that the UV image reductions here are optimised for a full field, which gives us slightly different integration times compared to the GASP jellyfish UV analysis \citep{George_2018,Poggianti_2019,George_2023}. The images from many orbits were co-added to create a master image for each cluster field. Astrometric calibrations were performed using the {\tt astrometry.net} package, in which solutions are performed using the USNO-B astrometric catalog \citep{Lang_2010}. Photometric calibrations were performed using zero-point values generated for photometric calibration stars, as is described in \citet{Tandon_2017} and updated in \citet{Tandon_2020}.

UVIT delivers UV images with an angular resolution of $\sim$1\farcs2 in the NUV channel\footnote{The NUV channel stopped functioning after two years of operation \citep{Ghosh_2021}.} and $\sim$ 1\farcs4 in the FUV channel \citep{Agrawal_2006,Tandon_2017}. The UV imaging of six clusters covers different regions with respect to their cluster centres. All six clusters were selected, in part, for the presence of one striking jellyfish galaxy, which we endeavoured to place at the centre of each telescope pointing. However, due to constraints from nearby bright objects (which, if violated, can suspend any ongoing observations and automatically switch off UVIT's three detectors), we slightly offset some fields to avoid such warnings. Thus, the position of the cluster
centre in the different images can change depending on the position of the GASP jellyfish galaxy and locations of nearby, UV-bright stars. Field locations, choice of filters, integration times, and other details of these FUV and NUV observations are presented in Table \ref{T1}.\footnote{PI: Koshy George, Proposal ID: G06${\_}$019 Abell 85, G07${\_}$002 Abell 1991, Abell 2626 G08${\_}$002 Abell 1668, A05${\_}$108 Abell 4059 PI: Bianca Poggianti A05${\_}$085 Abell 3376.}

Using these UVIT images, we carried out a search for new ram-pressure stripping galaxy candidates in these six galaxy clusters. Three individuals from the GASP team (BP, AO, and KG) visually inspected the NUV and FUV imaging for each of the six fields. We searched for galaxies that show evidence suggestive of ram-pressure stripping. These can be in the form of debris trails, tails, or debris located on one side of the galaxy and/or
asymmetric or disturbed morphologies suggestive of unilateral external forces, and/or a distribution of star-forming regions and knots suggestive of triggered star formation on one side of
the galaxy. These criteria are similar to those used in previous studies of jellyfish galaxies \citep{Ebeling_2014,Poggianti_2016}. We inspected all galaxies in the UV images of the cluster fields. The images were first inspected independently by each
classifier, who assigned a class according to the scheme
described below. The information from three classifiers was combined and the galaxy with the largest vote was selected as the candidate galaxy with the same procedure followed for the UV class.  The different classifiers agreed to within one
class in 75\% of the cases. Then, each galaxy was inspected
together by the common classifier to ensure homogeneity, the
final classification was agreed upon, and a consensus was found
on the classification of those galaxies whose individual class
differed by more than one class. We assigned our candidates to five classes
according to the visual evidence for stripping signatures in
the UV bands (UV class), from extreme cases (UV class 5) to
progressively weaker cases, down to the weakest (UV class 1). As
a result, our UV class=5 and 4 classes comprise the most secure
candidates with the most striking cases of ram-pressure stripping. UV class 3 candidates are probable cases of stripping, while UV classes 2 and 1 are
tentative candidates for which definitive conclusions cannot
be reached on ram-pressure stripping on the basis of the UV imaging. These systems contain unilateral features indicative of stripping but also contain evidence of tidal interaction on the galaxy disc. The UV class 2 and 1 candidates could therefore be also perturbed systems from tidal interaction, which we cannot rule out based on the evidence seen in UV imaging. Follow-up deep UV observations along with spatially resolved spectroscopic information from the ground are required to understand the true stripping nature of galaxies of UV classes 2 and 1. Information for the candidate jellyfish galaxies is presented in Table \ref{T3}. We also comment on the possibility of a tidal interaction for some of the galaxies.

 Three of our target clusters have both NUV and FUV images, while only FUV imaging data is available in three clusters. UVIT has an option to select from a set of broad and narrow band filters in NUV and FUV. We tried to use the same broad-band filters whenever possible, except for Abell 1668, for which imaging was performed using an FUV filter with a slightly narrower passband (i.e. F154W; see Table~1). Since the NUV images have slightly better spatial resolution and sensitivity compared to the FUV, we used NUV images, when available, for our analysis (Abell 1991, Abell 2626 and Abell 85); otherwise, we used only the FUV images (i.e. Abell 1668, Abell 3376, and Abell 4059).

\section{Results}

We identified 54 galaxies as stripping candidates based on unilateral emission apparent in their UV morphologies. These galaxies could be experiencing star formation within the gas stripped from the galaxy by cluster infall and subsequent interactions between the galaxy and the intra-cluster medium.
We performed an initial crossmatch of these galaxies with the photometric and spectroscopic catalogs from the WINGS/OmegaWINGS survey 
 \citep{Fasano_2006,Cava_2009,Gullieuszik_2015,Moretti_2017}. We found that, out of the 54 candidates, 51 returned a match with the photometric catalog; 39 also have a match in the spectroscopic catalog (within a matching radius of 5\arcsec). Of these, 19 have been spectroscopically confirmed as members, with velocities relative to the cluster median within three times the velocity dispersion of the cluster. 

\begin{figure}
{\includegraphics[width=0.5\textwidth]{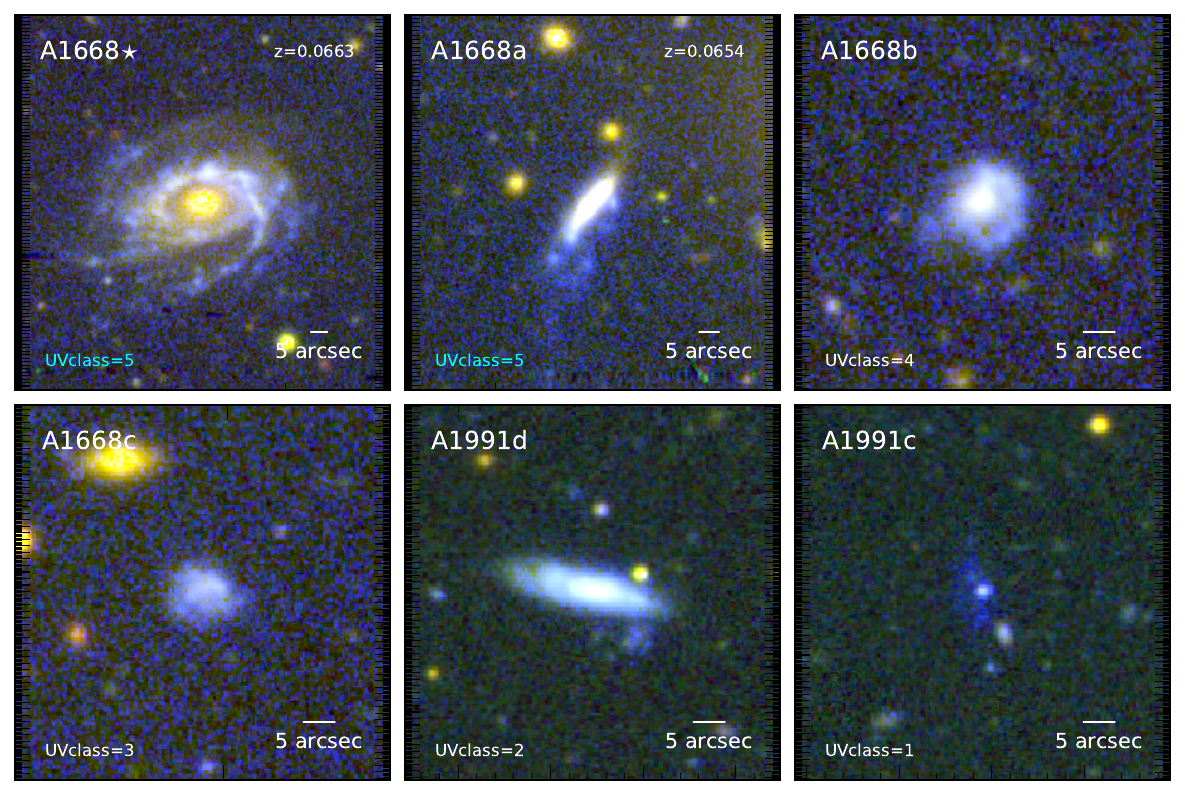}}
\caption{Colour composite (FUV,B,V) images of jellyfish galaxies with UV classes 5-to-1 from two clusters A1668 and A1991. The UV class assigned to the galaxy is noted, and the cyan colour denotes the confirmed spectroscopic cluster members.}\label{figure:A1668A1991RGB1}
\end{figure}

Table \ref{T2} gives details on the clusters, the region covered by UVIT at the cluster rest frame, the depth of the UV imaging data used for analysis, and the number of jellyfish galaxy candidates with and without spectroscopy for each cluster field.
We mark the jellyfish galaxy candidates as circles, along with the tail directions shown by an arrow, given in Zenodo. Galaxies that have spectroscopic redshifts are marked with a dashed circle, with cluster members shown in cyan. In each case, the jellyfish galaxy first identified in GASP is indicated with a larger cyan circle. The centre of each cluster is shown by a green cross. As an example, in Fig \ref{figure:A1668A1991RGB1} we present the colour composite images of galaxies created using FUV, B, and V with a UV class from 5 to 1, which illustrate the jellyfish nature of the candidate galaxies.

\begin{figure*}
\sidecaption
  \includegraphics[width=12cm]{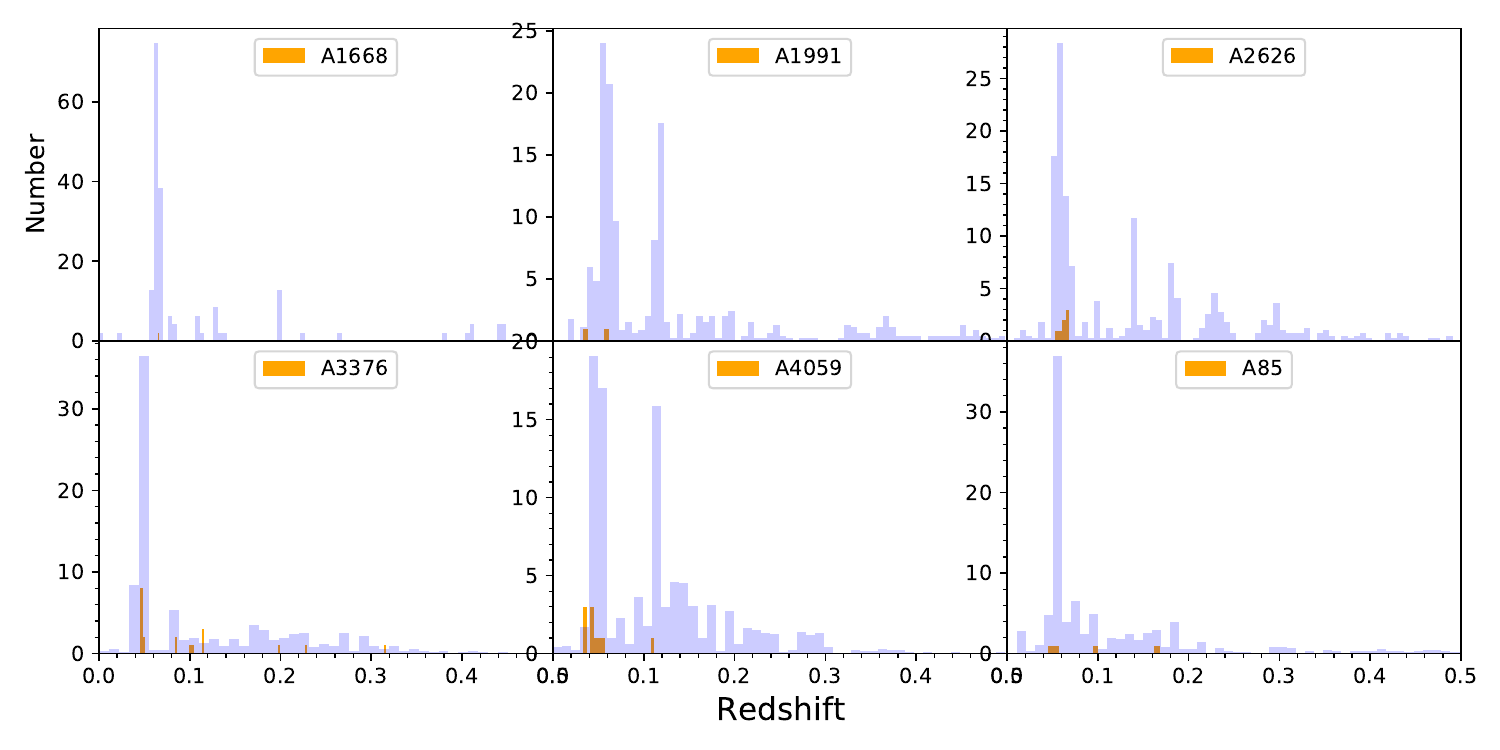}
     \caption{Redshift distribution of galaxies that have measured redshifts in each of our six clusters fields (blue histograms). The distribution of  jellyfish candidates from our analysis is shown in orange.}
     \label{figure:hist_z}
\end{figure*}

\begin{figure}
\centerline{\includegraphics[width=0.5\textwidth]{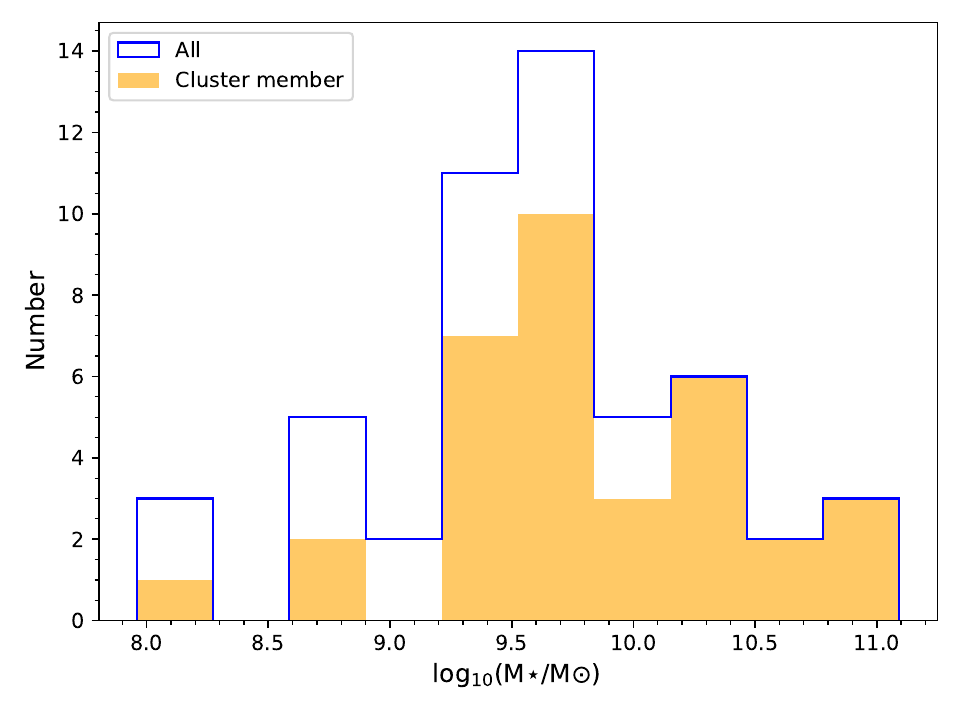}}
\caption{Stellar mass distribution of all jellyfish candidates from this analysis, in blue. For comparison, the distribution of candidates with spectroscopic redshifts that match their associated clusters is shown in orange.}\label{figure:hist_mstar}
\end{figure}

The spectroscopic redshift distribution of candidate jellyfish galaxies for all six clusters is shown in Fig \ref{figure:hist_z}. Spectroscopically confirmed members are plotted in orange, while other galaxies along the line of sight are shown in blue. There are galaxies in the foreground and background of the cluster, falling outside our requirement for membership in the cluster.
We checked the possibility that non-member jellyfish candidates identified from UVIT belong to foreground or background structures along the line of sight (e.g. other galaxy groups or clusters). In fact, the non-cluster members that have spectroscopic redshifts in five clusters\footnote{In Abell 1668, there is no spectroscopically confirmed, non-member UV candidate.} appear to be within structures at different redshifts, identifiable as additional peaks in the redshift distributions from the WINGS/OmegaWINGS catalogues. Unfortunately, due to target selection for WINGS-OmegaWINGS spectroscopy, we cannot quantify how massive these additional structures are from the relative importance of the redshift distribution peaks. However, the fact that these galaxies fall in distinct peaks is an indication that they might be ram-pressure stripped or disturbed galaxies in other structures along the line of sight. 
We checked in detail for the association of the non-cluster member candidates with known galaxy groups in the literature. Abell 2626 hosts six substructures, as is discussed in detail in \citet{Healy_2021}. We found that out of the three non-cluster member candidate galaxies in the Abell 2626 field, A2626c belongs to a substructure called ‘The Swarm’ (see Fig 10 of \citet{Healy_2021}). The Abell 4059 cluster field shows a substructure in the OmegaWINGS catalog. We did not find any of the four non-cluster members as part of this substructure.
For the other four clusters, we could not find any additional structures in the WINGS  catalogs. We searched for the presence of possible groups along the line of sight for all six cluster fields studied here by using the group catalog of \citet{Tempel_2014} based on Sloan Digital Sky Survey data release 10, which contains information on galaxies in groups up to a redshift of 0.2. We could identify one non-cluster member in the Abell 1991 field (A1991d) as part of a foreground group at a redshift of $\sim$ 0.0335.
The stellar masses (assuming a Chabrier IMF) for the candidate jellyfish galaxies were derived using the prescriptions of \citet{Bell_2001} for the spectroscopic redshift of the galaxies (or the redshift of the cluster when there is no available redshift measurement)  \citep{Vulcani_2022}. Their distribution is shown in Fig \ref{figure:hist_mstar}. We did not see any correlation between the stellar mass and the UV class assigned to the candidates.

We have identified 54 UV jellyfish candidates in the current study from our search using the UV imaging of six clusters. Out of these 54 candidates, seven have already been identified from previous optical B-band imaging in P16. There are then 47 new candidates from UV. Out of these 47, only 45 fall within the previous optical search footprint. We have two galaxies outside the previous optical search. We explored the reasons for not identifying the 45 galaxies as candidate galaxies based on our optical imaging, by comparing the optical B-band image with the UV images, and concluded, not surprisingly, that UV emission is more sensitive to recent star formation (100-200 Myr), and hence to low surface brightness extra-planar emission, than the B-band imaging is. Five striking examples are shown in Figs B.4 and B.5. These galaxies (A3376a, A4059a, A4059f, A4059h, and A4059i), missing from the P16 catalog, have UV classes of 5 and 4. A careful analysis reveals that two of these galaxies (A4059i and A4059f) fall into the inter-chip regions of the P16 imaging, while for the others the contrast
in UV imaging is better than the optical B-band in allowing us to classify them as ram-pressure stripped. For example, A3376a has clear extra-planar UV emission towards the north-east direction that is not detected in B-band imaging.

\begin{table*}
\caption{\label{T2} Basic information of six galaxy clusters, UVIT imaging, and number of identified jellyfish candidates.}
\centering
\label{galaxy cluster details}
\tabcolsep=0.1cm
\begin{tabular}{lcccccccrrr} 
\hline 
\hline
Galaxy cluster & $\langle{z}\rangle$ & $\sigma$ & R$_{200}$ & M$_{200}$ & R$_{uvit}$  & $\delta_{center}$ &  Image depth  & N$\rm cand$ & N$\rm spec$ & N$\rm mem$\\
&  &(km/s) & (Mpc) & (10$^{14}$ M$\odot$) & (Mpc)  & (Mpc) &  (mag for S/N=5)  & &  & \\

~~~~~~~ (1) & (2) & (3) & (4) & (5) & (6) & (7) & (8) & (9) & (10) & (11)\\
\hline
Abell 1668 & 0.0634  &   649$\pm$57  & 1.35 & 2.99   & 1.02  & 0.483  & 24.91$^{FUV}$ & 3    &   1   &  1     \\
Abell 1991 & 0.0584  &   599$\pm$57  & 1.33  & 2.83  &  0.95 & 0.610  & 26.51 $^{NUV}$ &4    &   1   &  0   \\
Abell 2626 & 0.0548  &   625$\pm$62  & 1.48 & 3.88   & 0.90  & 0.000  & 26.28$^{NUV}$& 6    &   6   &  3  \\
Abell 3376 & 0.0461	 &   779$\pm$49  & 1.65 & 5.34   & 0.76  & 0.489  & 25.85$^{FUV}$ & 21  &  19  &  9    \\
Abell 4059 & 0.0480	 &   715$\pm$59  & 1.58 & 4.69   &  0.79 & 0.339  & 26.04$^{FUV}$ & 10  &  9  &   5   \\
Abell 85     & 0.0521	 &  1052$\pm$68  & 2.02 & 9.88  & 0.85  & 0.000  & 26.92$^{NUV}$ &10  &   3   &   1  \\
\hline
\end{tabular}
\tablefoot{\scriptsize Key to columns: (1) cluster name; (2) mean redshift; (3) velocity dispersion; (4) the radius delimiting a sphere with an interior mean density
200 times the critical density; (5) mass of the cluster contained in R$_{200}$; (6) radius of the region covered at the cluster rest frame by the UVIT circular field of view; (7) offset between the cluster centre and the pointing centre of UVIT field; (8) image depth measured as the magnitude of a source with S/N=5 from the FUV or NUV image of the cluster field that is used for identification of stripping candidates;  (9) number of jellyfish candidate galaxies; (10) number of jellyfish candidates with spectroscopic redshifts; (11) number of jellyfish candidates that are confirmed cluster members.}
\end{table*}

\subsection{Individual clusters}

We now present details on the UV imaging of each of the six clusters. For each program field, the galaxy cluster centre is taken to be the position of the brightest cluster galaxy measured in the optical imaging. These positions match, in all cases, the cluster centres measured from X-ray imaging. For Abell 2626 and Abell 85, the UVIT imaging was carried out with the cluster centre coincident with the field centre. We indicate the cluster centres with a green cross in the cluster images. 
The tail directions measured for each of the GASP jellyfish galaxies (shown in cyan) are also used in our analysis. We now comment briefly on each of the targeted clusters and their candidate galaxies.

\begin{enumerate}

\item[]{\bf Abell 1668.}
The centre of Abell 1668 is offset by $\sim$0\fdg11 in UVIT imaging or $\sim$483 kpc in the cluster rest frame.  There are three new jellyfish galaxy candidates (in addition to the GASP JW39) identified in this cluster from UV imaging. One of these is a confirmed cluster member based on a spectroscopic crossmatch.
 The individual galaxies are shown in Fig \ref{figure:A1668RGB}.  A1668a is a spectroscopically confirmed cluster member that displays quite convincing 
 unilateral features typical of ram-pressure stripped galaxies. 

\item[]{\bf Abell 1991.}
The centre of Abell 1991 is offset by $\sim$0\fdg15 in UVIT imaging or $\sim$610 kpc in the cluster rest frame. There are four new candidates (in addition to GASP JO60) identified from visual inspection with only one galaxy (A1991d) that has spectroscopic redshift. The galaxy is not a cluster member but could be part of a foreground structure at $z$=0.0328. The individual galaxies are shown in Fig \ref{figure:A1991RGB}.

\item[]{\bf Abell 2626.}
The centre of Abell 2626 coincides with the centre of the UVIT imaging. There are six candidates (in addition to GASP JW100) identified from visual inspection, with three galaxies that are cluster members (Fig \ref{figure:A2626RGB}a, \ref{figure:A2626RGB}b, and \ref{figure:A2626RGB}f) based on the available spectroscopy. The individual galaxies are shown in Fig \ref{figure:A2626RGB}. 
Abell 2626b (Fig \ref{figure:A2626RGB}b) and  Abell 2626c (Fig \ref{figure:A2626RGB}c)  both show very convincing unilateral features typical of ram-pressure stripped galaxies.

\item[]{\bf Abell 3376.}
The centre of Abell 3376 is offset by $\sim$0\fdg15 in UVIT imaging or $\sim$489 kpc in the cluster rest frame. There are 22  candidates identified from visual inspection, with nine cluster members out of 19 galaxies having spectroscopic redshifts. The individual galaxies are shown in Fig \ref{figure:A3376RGB}. A low-surface-brightness artefact appears at the top left of the UVIT field of view, which could be due to the reflection of bright sources outside of the field from satellite components (but not affecting our analysis). The GASP jellyfish galaxy JW108 in this cluster, seen edge on, has a truncated disc in the UV and does not show emission from a stripped tail, which is otherwise seen in H$\alpha$ imaging from MUSE. In other words, this galaxy is not classified as a jellyfish from UV imaging. Galaxies A3376g and A3376m have HI and H$_2$ detection from MeerKAT and APEX observations. A3376g has a HI mass of 10$^{8.8}$ M$\odot$ and a H$_2$ mass of 10$^{9}$ M$\odot$. A3376m has a HI mass of 10$^{9}$ M$\odot$ and a H$_2$ mass of 1.6 $\times$ 10$^{9}$ M$\odot$ \citep{Moretti_2023}.

\item[]{\bf Abell 4059.}
The centre of Abell 4059 is offset by $\sim$ 0\fdg10 in UVIT imaging or $\sim$339 kpc in the cluster rest frame. There are ten candidates (in addition to GASP JO194)  identified from visual inspection, with five cluster members out of nine galaxies that have spectroscopic redshifts. The individual galaxies are shown in Fig \ref{figure:A4059RGB}.  Fig \ref{figure:A4059RGB}a, f, g, i, and j  are cluster members based on available spectroscopic redshift information. Fig \ref{figure:A4059RGB}a appears to be a prominent example of a strong ram-pressure stripping galaxy with unwinding spiral arms similar to the cases discussed in \citet{Bellhouse_2021}.

\item[]{\bf Abell 85.}
The centre of Abell 85 coincides with the centre of the UVIT imaging. There are ten candidates (in addition to GASP JO201)  identified from visual inspection with one cluster member out of three galaxies having spectroscopic redshifts.  The individual galaxies are shown in Fig \ref{figure:A85RGB}.  Fig \ref{figure:A85RGB}c is a spectroscopically confirmed member that may host prominent clumps of star formation in a stripped tail. 
\end{enumerate}

\subsection{Tail directions}

\begin{figure}
\centerline{\includegraphics[width=0.5\textwidth]{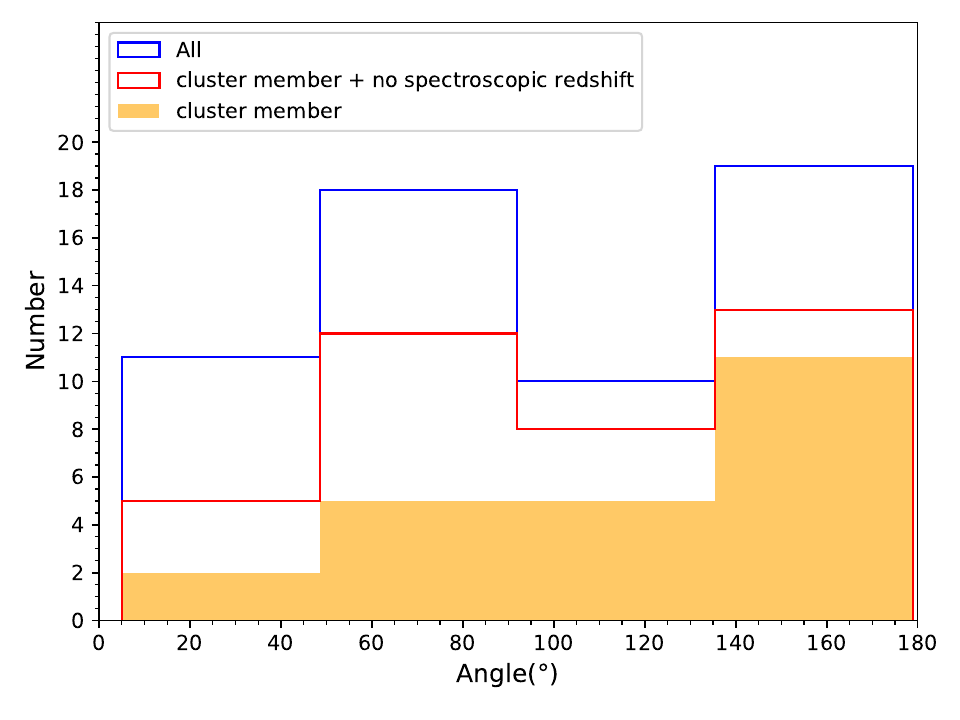}}
\caption{Distribution of the angle between the direction of the tail and the direction towards the cluster centre. The tail angle distribution of all jellyfish candidates is given in blue. The distribution of candidates with a spectroscopic redshift matching with the associated cluster membership is shown in orange. The combined distribution of galaxies with no spectroscopic redshift and confirmed cluster membership are shown in red. We note that there are two peaks in distribution at  70$\degree$ and 160$\degree$ when we include all the candidates and also for galaxies with no redshift information.}\label{figure:jf_tailangle_hist}
\end{figure}

The orientations of the stripped tails of candidate jellyfish galaxies with respect to the cluster centre can provide constraints on the orbital dynamics of galaxy since infall \citep{Rory_2022,Salinas_2024}. First-infall galaxies undergoing ram-pressure stripping should have tails preferentially pointing away from the cluster centre, whereas those observed after a pericentre passage can exhibit tails pointing towards the cluster centre. We estimated the projected tail direction with respect to the centre of the galaxy described in \citet{Roberts_2020}. The tail angle changes clockwise from 0 to 360$\degree$, placing the galaxy at the centre. We then computed the shortest angle between the tail direction and the centre of the galaxy cluster. For galaxies with a projected tail direction oriented towards the galaxy cluster centre, this angle will be around 0$\degree$, whereas it will be 180$\degree$ for tails oriented away from the cluster centre. We include here the four GASP jellyfish galaxies with a tail direction that can be measured from UV imaging in this analysis (JO201, JW100, JW39, and JO194). In all, there are 58 galaxies with estimated tail directions. The distribution of the measured angle for these 58 galaxies is shown in blue in Fig \ref{figure:jf_tailangle_hist}. Following \citet{Rory_2022}, we used wide angle bins 45$\degree$ in size to ensure that the distribution is not affected by any small inaccuracies in the measurement of the tail angle. For comparison, the distribution of the 23 confirmed cluster members is shown in orange.  There are two peaks in the distribution at around 70$\degree$ and 160$\degree$ with all galaxies, which disappears when only the confirmed cluster members are included. We also show the combined distribution of 15 candidates with no redshift information along with the 23 confirmed cluster members. The double peak that we see here could be due to the inclusion of  galaxies that are not part of the cluster. This could explain the disappearance of such a behaviour when we include only the confirmed cluster members. Recent simulation-based studies of the gaseous tails of jellyfish galaxies have found that the tails extend in opposite
directions to the galaxy trajectory, with no relation between tail orientation and the position of the galaxy cluster's centre (see Fig 7 of \citet{Yun_2019}). We have found that tails are oriented away from the cluster centre for confirmed cluster members. Our tail orientation results for confirmed cluster member candidates are consistent with the results drawn for the jellyfish galaxies' tail orientation in the Coma cluster (see Fig 4 of  \citet{Roberts_2020}).

We also note that the dynamical state of galaxy clusters can disturb the tail direction of the ram-pressure stripping fraction of galaxies \citep{Salinas_2024}. If there is a massive substructure in the cluster and the galaxy interacts with the substructure, the tail direction may not be measured with respect to the cluster centre. The Abell 2626, Abell 3376, Abell 4059, and Abell 85 clusters are known to have signatures of recent interactions \citep{Ana_2023}. We did not see any correlation between the tail angle and the UV class assigned to the candidates with confirmed cluster membership. There are 11 galaxies of UV class 4 and 5, of which six show a tail angle of 180$\degree$. There are 12 galaxies of UV class 1 and 2, of which five show a tail angle of 180$\degree$. We remind the reader that the UVIT imaging has a 28$\arcmin$ diameter field of view and the observations on which this study is based were optimised for six GASP jellyfish galaxies in six clusters. We tried to keep the GASP jellyfish galaxy at the centre of the UV imaging. The details of the region covered by UVIT imaging at the cluster rest frame and the offset between the cluster centre and UVIT pointing are given in Table \ref{T2}. We do not cover regions uniformly with respect to the cluster centre in the UV imaging of the six clusters. The six clusters' FUV/NUV imaging was performed with different integration times. We note the corresponding magnitude of a source with a signal-to-noise (S/N) of 5, measured from the FUV/NUV images used for the identification of the candidates from each cluster. The difference in the image depths between the cluster fields can introduce biases in the detection of low-surface-brightness features, leading to the identification of ram-pressure stripped candidates within individual clusters.\\

\subsection{Phase-space diagram}

\begin{figure}
\centerline{\includegraphics[width=0.5\textwidth]{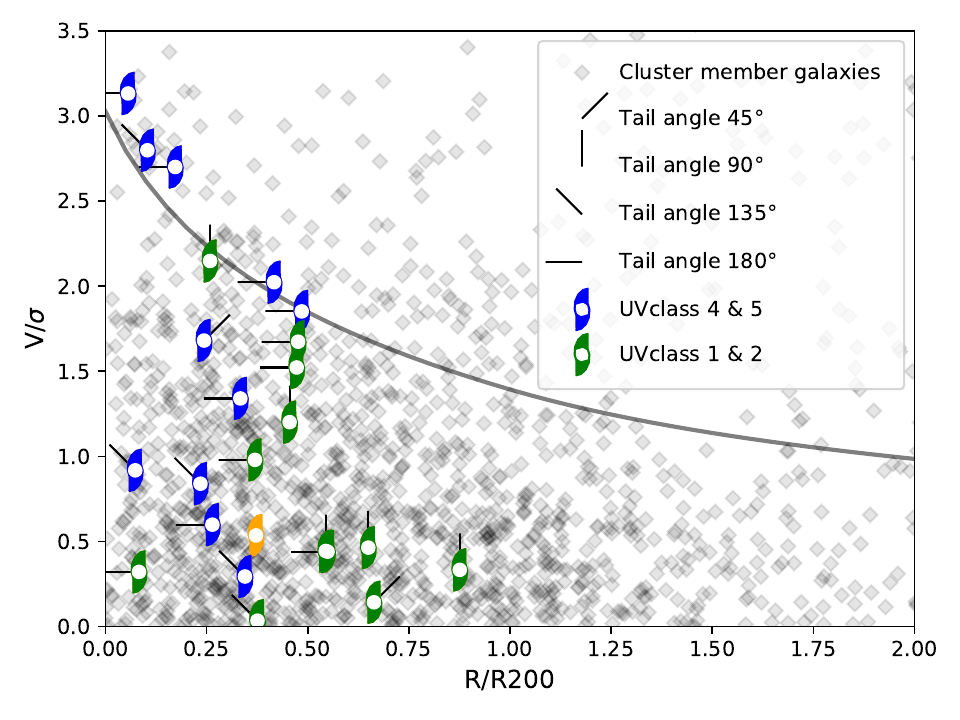}}
\caption{Phase-space diagram for member galaxies in our six program clusters (grey diamonds). We assign different-coloured spiral-shaped points for the UV class of the jellyfish galaxy candidates from our analysis of UVIT imaging. We assign the UV class as two groups in the colour scheme here to examine the collective behaviour in the plot. The tail angle of these galaxies is divided into four groups of 45$\degree$ each and each group is assigned a line of orientation that changes progressively with the group. Note that we plot the normalised, projected distance from the cluster centre (in units of R200) against the galaxy line-of-sight velocities with respect to the cluster mean, also normalised to cluster velocity dispersion. The three-dimensional (unprojected) escape velocity in an NFW halo with a concentration of $c $= 6 is shown by the grey curve. Galaxy JW108 from the A3376 cluster, which has a very pronounced truncation of star formation within its disc, is shown in orange.}\label{figure:vbsigma}
\end{figure}

The orbital histories of jellyfish candidates with confirmed cluster membership were investigated using the location of galaxies in the projected position versus radial velocity phase-space diagram. Fig \ref{figure:vbsigma} shows the cluster-centric distance versus $\Delta v_{cl}$/ $\sigma_{cl}$, the ratio of the differential velocity of the galaxy and the cluster velocity dispersion. The distances have been normalised by R200, the radius within which the mean density is 200 times the critical density of the Universe (which approximately corresponds to the cluster virial radius). Galaxies with low velocities with respect to the cluster (i.e. $\Delta v_{cl}$/$\sigma_{cl}$ $<1$ ) are expected to have completed at least one cluster crossing, whereas galaxies on the first infall are preferentially found with higher velocities over a range of cluster radii (see e.g. Fig. 6 of \citet{Jaffe_2018,Jaffe_2019} for a visualisation of the stripping sequence in regions of the phase-space based on modelling).  The candidate galaxies are all found close to the centre in the phase-space diagram because that is the region covered by UVIT. The confirmed cluster member jellyfish candidates cover a range of velocities and cluster centric radii within 1 $\times$ R200. 
 We note that the jellyfish candidates shown in Fig. \ref{figure:vbsigma} are mostly within 0.5 $\times$ R200 but their radial distribution in this diagram is affected by incompleteness beyond a radius that varies from cluster to cluster, depending also on the region covered by the UVIT pointing at the cluster redshift. This implies that the appearance of galaxies in the ‘virialised region’ does not mean they are virialised and the information on the true distribution of galaxies in phase-space is missing. We might be able to partially differentiate between galaxies undergoing first infall and galaxies that complete first pericentre crossing by looking at their vertical position in phase-space but it is quite difficult to detect the difference between the two populations at a small projected radius.

The other cluster non-member candidate galaxies with spectroscopy have high $\Delta v_{cl}$/ $\sigma_{cl}$ ($>$ 4) values. This is, of course, expected as they are not part of these six clusters but lie in other structures along the line of sight. There are 15 candidate galaxies that have no spectroscopic information and that hence could not be included in the phase-space diagram analysis.

\section{Discussion}

The galaxy cluster velocity dispersion measured from the member galaxies' line-of-sight velocities is an indicator of the gravitation potential. The six galaxy clusters presented here show a range of velocity dispersion, and there is no apparent trend in the number of new candidates identified with cluster velocity dispersion. This is in line with the studies based on cluster galaxies that show signs of stripping in optical imaging \citep{Vulcani_2022}.  Abell 85 has a velocity dispersion comparable to the most massive galaxy clusters in the Universe but we find relatively few new stripping candidates from UVIT imaging. Abell 3376, on the other hand, has the largest number of confirmed member candidates identified and is the cluster with the longest integration time. The stripped tails in this cluster exhibit knots of star formation as well as diffuse emission in the UV ---  low-surface-brightness features that may only be evident in deep integrations.

The well-studied ‘benchmark’ galaxy cluster in the local Universe, Coma, was recently surveyed for ram-pressure stripping candidates by \citet{Roberts_2020}. These investigators visually inspected colour composite $u,g,i$  images with a resolution of $<$ 1$\arcsec$ taken with the Canada-France-Hawaii telescope and covering the region inside the virial radius ($\sim$ 9 deg$^2$) a total of 41 candidates were found. The tail directions of these candidates were measured in a manner similar to that used here, with a majority of angles falling between 120 and 180 \degree. It was also found that the tails of Coma cluster ram-pressure stripped candidates are often oriented away from the cluster centre, as would be expected for infalling galaxies. This preferential orientation in Coma was also noted by \citet{Smith_2010}, based on GALEX UV imaging, and by \citet{Yagi_2010} from H$\alpha$ imaging.  Interestingly, the H$\alpha$ and broad-band optical searches, even  within the same regions, yielded slightly different samples. This show again that different tracers or wavelength regions are sensitive to different characteristics, including differences in the stage of the stripping event. 

Unlike these studies of the Coma cluster, this study does not cover the virial radius uniformly for the six clusters in our sample, which makes a direct comparison of results problematic. Also, UV and optical imaging probe star formation at different timescales, which could lead to different sensitivities for detecting jellyfish candidates. The six clusters in our survey are located at  higher redshifts --- 0.046 $\lesssim z \lesssim$ 0.063, compared to $z =$ 0.0233 for Coma --- and are imaged in UV with better 1\farcs2--1\farcs4 angular resolution ($\sim$1.3 kpc at median z $\sim$ 0.048 ). The previous studies of Coma cluster jellyfish candidate galaxies were carried out using GALEX, which has an angular resolution of 4\farcs3--5\farcs3 ($\sim$2 kpc at z $\sim$ 0.0233). 
An important caveat, however, is the difference in image depths among our program clusters (see Table~2). The Abell 4059 and Abell 3376 observations had the longest integration times and, perhaps not surprisingly, produced the largest number of jellyfish candidates. Some additional cluster-to-cluster differences may be expected given that different wavelengths were used in the search and that only three clusters had available NUV imaging. Because the NUV filter response is almost twice that of the FUV, one would need deeper integrations to reach the same sensitivity level as in the NUV (see Fig 1 of \citet{Tandon_2017}).

These issues aside, we identified 54 ram-pressure stripping candidates across six galaxy clusters. The galaxies cover a range of stellar masses, as is shown in Fig \ref{figure:hist_mstar}. There are 39 galaxies with spectroscopic redshift information, of which we find 19 that are spectroscopically confirmed members. This suggests a success rate of $\sim$50 $\%$ for establishing cluster membership. But one question is whether the other candidates could be genuine stripping candidates, just not belonging to the target cluster. In the case of Abell 2626, this is clearly demonstrated by spectroscopic observations for galaxies in a 2 $\times$ 2 deg$^2$ region centred on the galaxy cluster (see Fig 10 of \citep{Healy_2021}). There are two background structures that appear in this region at slightly higher redshifts with different positions on the phase-space diagram of the cluster \citep{Healy_2021}. One of the candidates in Abell 2626c is a non-member of Abell 2626 but part of the background structures (The Swarm). Additional, targeted studies of the candidate jellyfish galaxies will be needed to fully understand the nature of their disturbed UV morphologies.

The extent and orientation of jellyfish tails may vary depending on the criteria used in identifying the candidate galaxies. Emission that traces the hot ionised gas, like H$\alpha$ or the radio, can be more extended than emission from star formation alone, traced by UV imaging. There are individual jellyfish galaxies that exhibit a one-to-one correspondence between the presence of H$\alpha$ and UV emission \citep{George_2018}. There are also cases of no one-to-one correlation in the tails of jellyfish galaxies \citep{Poggianti_2019}. Also, there may be a positional offset due to different timescales involved in stars contributing to UV and H$\alpha$ emission. The stripped gas can decouple and continue to accelerate, leaving behind streams of newly formed stars that are no longer influenced by the intra-cluster medium \citep{Kenney_2014,Poggianti_2019,Giunchi_2023b}. This may be further affected by viewing geometry, in the sense that candidate galaxies moving along the line of sight will likely be missed and those moving with a small angle with respect to the line of sight will have a smaller tail compared to galaxies seen moving in the plane of the sky. Thus, the projected direction of the tail seen from the imaging data depends significantly on the viewing angle. Also, galaxies can have complicated tail directions, like the case of GASP jellyfish galaxy JO201 in Abell 85 with unwinding spiral arms. The unwinding is understood to happen due to differential ram pressure caused by the disc rotation, causing stripped material to fall to higher orbits \citep{Bellhouse_2021,Vulcani_2022}. We have two candidate galaxies, A4059a and A2626c, along with the GASP jellyfish galaxies JO201, JO194, JO60, and JW39 from our analysis, that appear to have unwinding spiral arms.

 A statistically large sample of candidate jellyfish galaxies from many clusters is thus required, at different wavelengths, to better understand the galaxy infall and stripping processes. This work is a first attempt at building such a sample at UV wavelengths for six galaxy clusters; it represents the largest sample of clusters with UV-detected jellyfish candidates to date. In building this sample, we have relied on visual methods to search for unilateral UV emission from candidate jellyfish galaxies. It is known that certain galaxies can undergo processes (such as tidal disruption or mergers) that give rise to features similar to those produced by ram-pressure stripping. A detailed follow up of the UVIT-selected candidates, using optical IFU observations (like GASP), could provide kinematic maps for confirming, or refuting, the true ram-pressure nature of these galaxies. Additional UV imaging to larger cluster centric radii for these clusters, or for clusters in different dynamical states and/or distances, would improve our understanding of ram-pressure stripping throughout the different phases of cluster assembly. We have FUV imaging data of 14 more clusters, from which we have identified new candidate ram-pressure galaxies. When combined with the current catalog of candidates reported in this paper, this can be used for a statistical analysis of the tail direction. We plan to combine the UV imaging data with other multi-wavelength data to make a rigorous analysis of the star formation history of the new ram-pressure stripped galaxies.

\section{Summary}

We have carried out wide-field NUV and FUV imaging to search for evidence of ram-pressure stripping of galaxies belonging to six galaxy clusters between $z\sim0.048$ and $z\sim0.063$. The UV imaging is highly sensitive to recent (100-200 Myr) or ongoing star formation in galaxies that show asymmetric, unilateral features indicative of such stripping events (jellyfish galaxies). Each of our program clusters was previously studied in the GASP survey and is known from optical imaging to contain a prominent jellyfish galaxy.
Using new FUV and/or NUV images from the UVIT instrument onboard Astrosat, we searched for new jellyfish candidates in these six galaxy clusters. We identified 54 candidate galaxies that show signs of unilateral extra-planar emission likely due to ram-pressure stripping. Seven of these have already been identified as likely stripping based on optical B-band imaging. We identified 47 new candidates from UV imaging. Spectroscopic redshift information is available for 39 of these candidate galaxies, of which 19 are associated with six clusters. The other 20 non-member galaxies are most likely stripping candidates from structures (i.e. galaxy groups or galaxy clusters) along the lines of sight for these clusters, which, in some cases, are confirmed from the existing spectroscopy.  We have measured and presented tail directions for the jellyfish galaxy candidates, which have a bimodal angle distribution, and examined the position of the candidate galaxies in the phase-space diagram. The tails are found to be oriented away from the cluster centre for confirmed cluster members. Though suffering from limitations created due to different integration times and also in imaging different regions with respect to the cluster centre, the current study reports the detected candidate ram-pressure stripped galaxies from six clusters, which will be extended to more clusters in a future work.

\section{Data availability}
 The UVIT image for five cluster fields can be accessed from Zenodo (https://doi.org/10.5281/zenodo.13743286).

\begin{acknowledgements}
This paper uses the data from the AstroSat mission of the Indian Space Research  Organisation  (ISRO),  archived  at  the  Indian  Space  Science  Data Centre (ISSDC).  
This project has received funding from the European Research Council (ERC) under the European Union's Horizon 2020 research and innovation programme (grant agreement GASP n. 833824. Y.J. acknowledges financial support from ANID BASAL project No. FB210003, FONDECYT Regular projects No. 1230441 and 1241426.)

\end{acknowledgements}


%
   
%


\begin{appendix}

\section{Table with details of candidate jellyfish galaxies.}

\begin{table*}
\caption{\label{T3}Details of candidate jellyfish galaxies in six galaxy clusters.}
\centering
\tabcolsep=0.1cm
\footnotesize
\begin{tabular}{cccccccc} 
\hline
Candidate & $z$ & Cluster Member & RA$_{J2000}$  & Dec$_{J2000}$ & log$_{10}$M$\star$/M$\odot$ & UV class & Notes and comments\\
 (1) & (2) & (3) & (4) & (5) & (6) & (7) & (8)  \\
 \hline
 A1668$\star$   &   0.0663  &        Y & 13:04:07.719  &  $+$19:12:38.41 & 11.17 & 5  & JW39\dag
\\
 A1668a        &  0.0654   &        Y & 13:03:52.39   &  $+$19:15:59.4   &   9.95 & 5  & \dag
\\
 A1668b        & --            &      --  & 13:04:33.57 & $+$19:26:22.0 &   9.35 & 4 & \dag
 \\
 A1668c        &  --          &       -- & 13:03:30.25 & $+$19:14:28.2 &  8.75  & 3 & \dag
\\
 A1991$\star$  &0.0622 & Y   & 14:53:51.567  &  $+$18:39:04.79  & 10.28   & 5 & JO60 \dag
 \\
 A1991a  &--  & --     & 14:53:13.77 & $+$18:46:09.03 &  --  & 3 &  \\
A1991b  &--   & --     &    14:53:47.05 & $+$18:38:52.4& 9.24    & 1 &  \\
A1991c   &--   & --     &   14:53:44.10 & $+$18:39:28.6& 8.13   & 1 &  \\
A1991d  & 0.0328 &  N & 14:54:46.55 & $+$18:41:38.4    & 9.27 &  2 &  \\
 A2626$\star$ & 0.0619  &  Y  & 23:36:25.054  &  $+$21:09:02.64 &  11.21 &  5 &  JW100, \dag
\\
 A2626a      &  0.0533  & Y &  23:36:18.59 & $+$21:04:02.7 &  10.07 & 4 &  \dag
\\
  A2626b   &   0.0611 & Y  &   23:36:26.54 & $+$21:10:54.7 &   9.72 & 5 &  \dag
\\
  A2626c  &   0.0659 & N &  23:35:32.78 & $+$21:10:11.1 &  9.67  & 3 &   \\
  A2626d  &   0.0681 &N &  23:36:09.46 & $+$21:11:10.6 &  9.85  & 3 &  \dag
 \\
  A2626e   &  0.0646 & N   & 23:35:57.10 & $+$21:11:06.1  & 10.41  &  1 &  \\
  A2626f    &  0.0584  & Y  & 23:37:12.76 & $+$21:04:06.5  &  9.61  & 1 &  \\
A3376$\star$ & 0.0479      & Y&  06:00:47.944 &   $-$39:55:06.90 &  10.37  & -- &  JW108\dag
\\  
 A3376a & 0.0480 & Y & 06:01:06.96 & $-$39:55:01.8 &  10.75  & 4  &  \\
  A3376b & 0.0845 & N & 06:01:38.70 & $-$39:55:06.6 &  9.31  & 1 &  interacting\\
  A3376c & 0.0834 & N & 06:00:59.69 & $-$39:57:59.7 &  9.69  & 1 &  \\
  A3376d  & 0.0490 & Y & 06:00:44.63 & $-$39:58:16.4 &  9.34  & 1 &  \\
  A3376e & 0.1151  & N & 06:00:57.54 & $-$40:00:28.8 &  9.63 & 1 & \\
 A3376f & --   & --  &  06:00:17.28 & $-$39:57:02.2 &  8.92  & 1 & interacting \\
 A3376g &  0.0469 & Y & 05:59:58.37 & $-$40:00:34.8 & 9.39  &  1 &  \\
 A3376h&--   & --  & 6:00:02.884 & $-$40:01:59.98 &  --  & 1 & \\
A3376i &  0.0507  & Y &  06:00:28.69 & $-$40:01:38.2 &  8.20  & 1 & \\
  A3376j &  0.3161 & N & 06:00:31.68 & $-$40:01:22.6 &  11.09  & 1 &  \\
  A3376k &  0.1980  & N  &  06:00:11.50 & $-$40:04:28.0 & 9.12  &  1 &  \\
  A3376l & 0.0476    & Y  &   06:00:07.20 & $-$40:06:14.4 &  9.31 &  1 &  \\
  A3376m &  0.0473  &Y & 06:01:42.91 & $-$39:56:39.7 & 10.24   & 2 &  \dag
\\
  A3376n  &  0.1155 & N & 06:01:31.78 & $-$39:49:59.0 & 10.21   & 1 &  \\
  A3376o  &  0.1004 & N & 06:00:40.49 &  $-$39:51:57.5 &  9.90  & 1 &  \\
  A3376p  & 0.1048 & N & 05:59:54.37 & $-$39:50:00.0  &  9.58  & 1 &  \\
  A3376q  &  0.2284 & N & 05:59:46.80 & $-$39:47:26.6 &  10.25  & 1 &  \\
  A3376r & 0.0476  & Y &  06:00:45.15 & $-$39:46:48.5 & 10.97   & 2 & ring \\
  A3376s &  0.1149 & N & 06:00:53.25 & $-$39:47:49.6 &  9.59 & 1 &  \\
  A3376t & 0.0454 & Y & 06:01:10.75 & $-$39:43:55.9  & 9.71   & 2 &  \\
  A3376u & 0.0473 & Y &  05:59:45.70 & $-$39:45:14.8 &  9.72 & 1 &  \\
  A4059$\star$ & 0.0420 &Y & 23:57:00.740  &   $-$34:40:49.94 & 11.58  &  4 &  JO194\dag
\\
  A4059a   &  0.0538 &  Y &  23:56:44.34 & $-$34:34:36.5 &   10.86 & 5 &  \\
  A4059b  &  0.0333 & N &  23:57:12.96 & $-$34:28:19.4 & 10.38   & 1 &  \\
  A4059c  &  0.0337 & N & 23:57:11.38 & $-$34:30:31.6 &   10.36 &  2&  possibly interacting \\
  A4059d   & 0.1116  &  N & 23:57:10.19 & $-$34:31:36.3 &  9.50  &  2&  \\
  A4059e   &  0.0336 & N & 23:56:12.45 & $-$34:31:55.5 &  8.62   & 2 &  \\
  A4059f  &  0.0446 & Y & 23:57:33.48 & $-$34:45:07.1 &  9.80  & 5 &  \\
  A4059g  &  0.0434 & Y &  23:57:32.71 & $-$34:48:26.6 & 8.76  & 2  &  \\
  A4059h  &--    &--   & 23:57:16.13 & $-$34:48:00.0 &   9.40 & 5 &  \\
  A4059i  &  0.0480  &  Y & 23:56:14.51 & $-$34:45:34.9 & 9.59  & 5  &  \\
  A4059j  & 0.0517& Y &  23:56:03.19 & $-$34:41:25.7 &  9.27  & 2 &  \\
 A85$\star$  & 0.056 & Y & 00:41:30.295  &   $-$09:15:45.98 & 10.81   & 5 &   JO201\dag
\\
  A85a   &--   &--   &  00:41:12.77 &  $-$09:12:41.9 &  7.96  & 1 &   \\
  A85b.    &--   &--  &  00:41:14.28 & $-$09:12:07.1 &  --  & 2 &  low surface brightness tail \\
  A85c    &  0.0556  & Y & 00:42:09.81 & $-$09:28:52.2 &  9.27 &  2 &  \\
  A85d    & 0.0953  & N & 00:42:11.21 & $-$09:28:20.0  &   9.76 & 1 &   \\
  A85e  &  0.1684  & N  & 00:42:14.98 & $-$09:28:28.1  & 10.54  &  1 &  \\
  A85f    &   --    &--   & 00:40:54.69 & $-$09:22:15.8. &  8.69  & 1 &   \\
  A85g     & --     & --  &  00:41:41.52 & $-$09:12:47.4 & 8.83   & 1 &  \\
  A85h  &  --    &--   &  00:41:15.28 & $-$09:20:23.7 &  10.04  & 2 &  \\
  A85i    &  --    &--   &  00:42:04.68 & $-$09:22:29.8  &  9.54  & 1 &  possibly interacting \\
  A85j  &  --     & --  &  00:42:02.36 & $-$09:23:03.4 &  9.56  & 1 &  \\
\hline
\end{tabular}
\tablefoot{\scriptsize Key to columns: (1) Candidate name assigned as alphabets added to cluster name. The GASP-confirmed jellyfish galaxy in each cluster is marked with a $\star$ with the corresponding name given in comments; (2) Redshift of candidate from \citet{Moretti_2014,Moretti_2017};  (3)  Galaxy cluster membership is shown with "Y" and known cases of non-members are marked with "N";  (4-5) equatorial coordinates of the galaxy centre; (6) logarithm of the galaxy stellar mass (in solar masses); (7) UV Jellyfish class. A class of 5 is the top score given to jellyfish secure galaxies, with classes of 4 to 1 decreasing in certainty; (8) Comments on candidate galaxies with stripped tails seen in NUV/FUV images. Galaxies previously identified as stripping candidates in P16 are marked with a \dag.}
\end{table*}

\section{Colour scale images of jellyfish galaxies}

\begin{figure*}
\centerline{\includegraphics[width=0.53\textwidth]{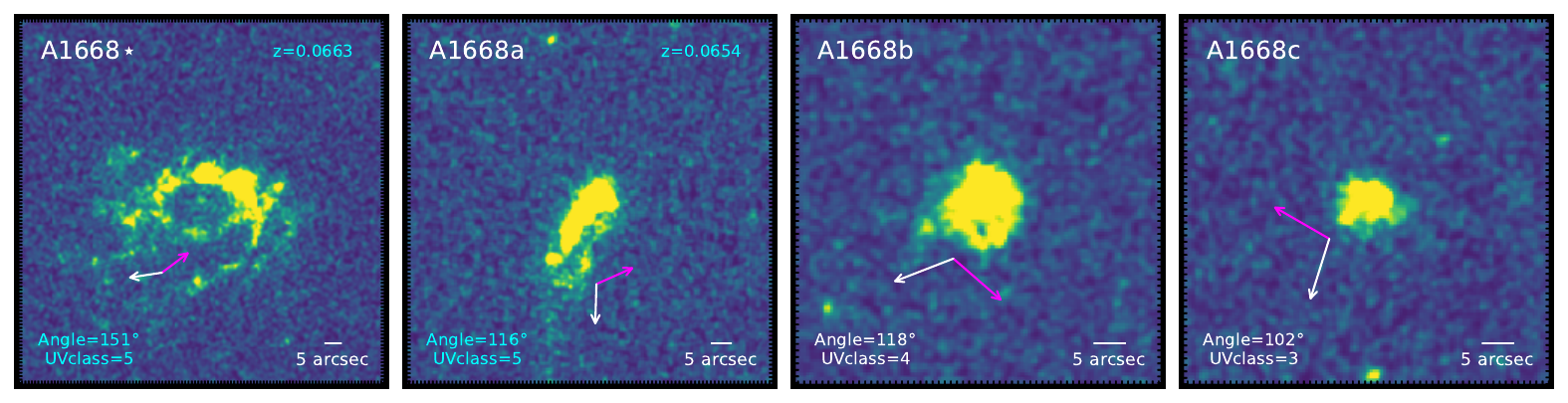}}
\caption{FUV colour scale images of 4 jellyfish galaxies belonging to Abell 1668 galaxy cluster. The 5$\arcsec$ scale bar shown corresponds to 6.1 $kpc$ at cluster rest frame. The UV class assigned to the galaxy is noted with text in cyan colour for the confirmed spectroscopic cluster members. The measured tail direction is shown with white arrow and the direction to cluster centre with magenta arrow. The angle between the tail direction and the cluster centre is given as tail angle.}\label{figure:A1668RGB}
\end{figure*}

\begin{figure*}
\centerline{\includegraphics[width=0.53\textwidth]{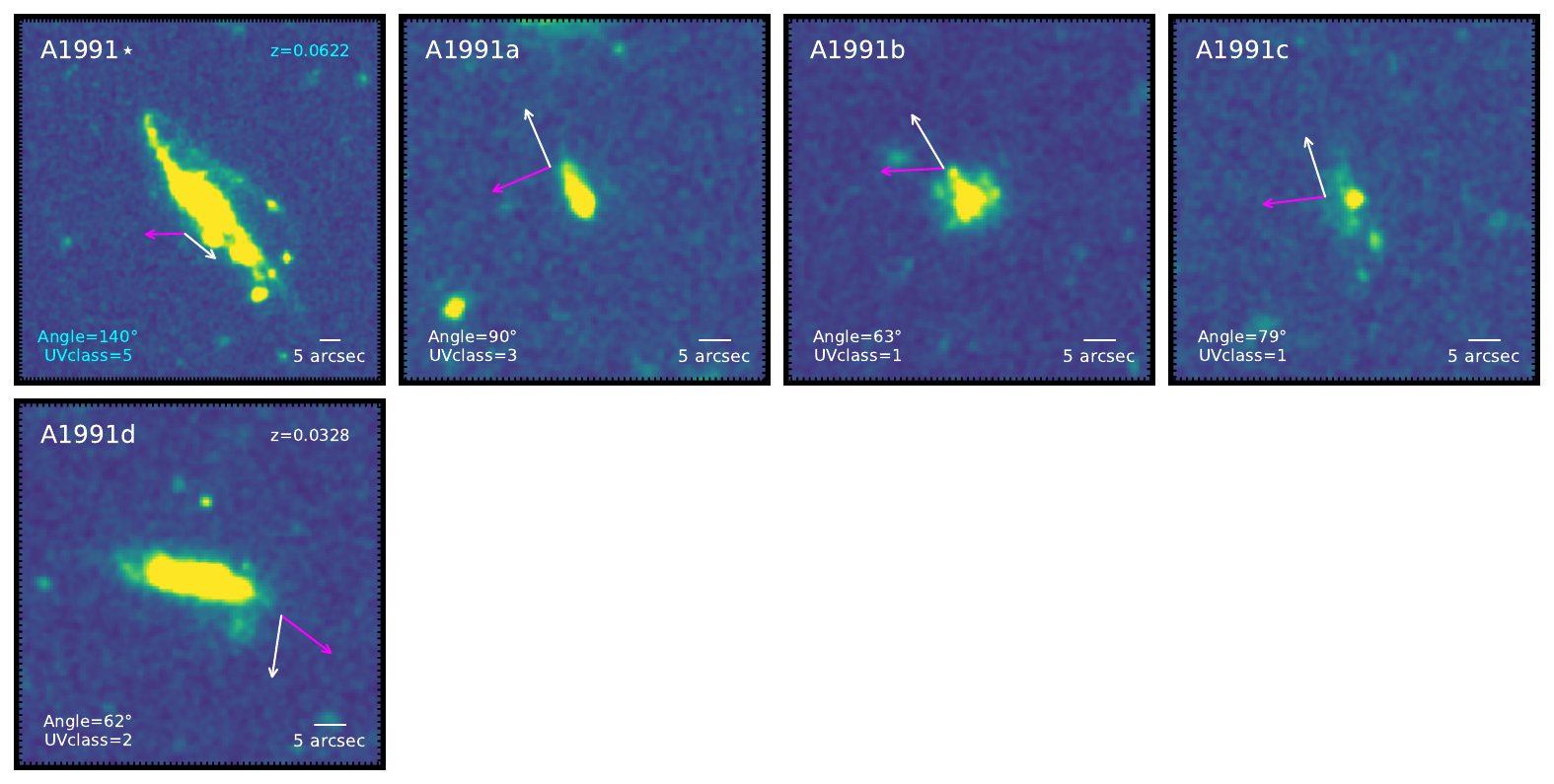}}
\caption{NUV colour scale images of 5 jellyfish galaxies belonging to Abell 1991 galaxy cluster. The 5$\arcsec$ scale bar shown corresponds to 5.65 $kpc$ at cluster rest frame. }\label{figure:A1991RGB}
\end{figure*}

\begin{figure*}
\centerline{\includegraphics[width=0.53\textwidth]{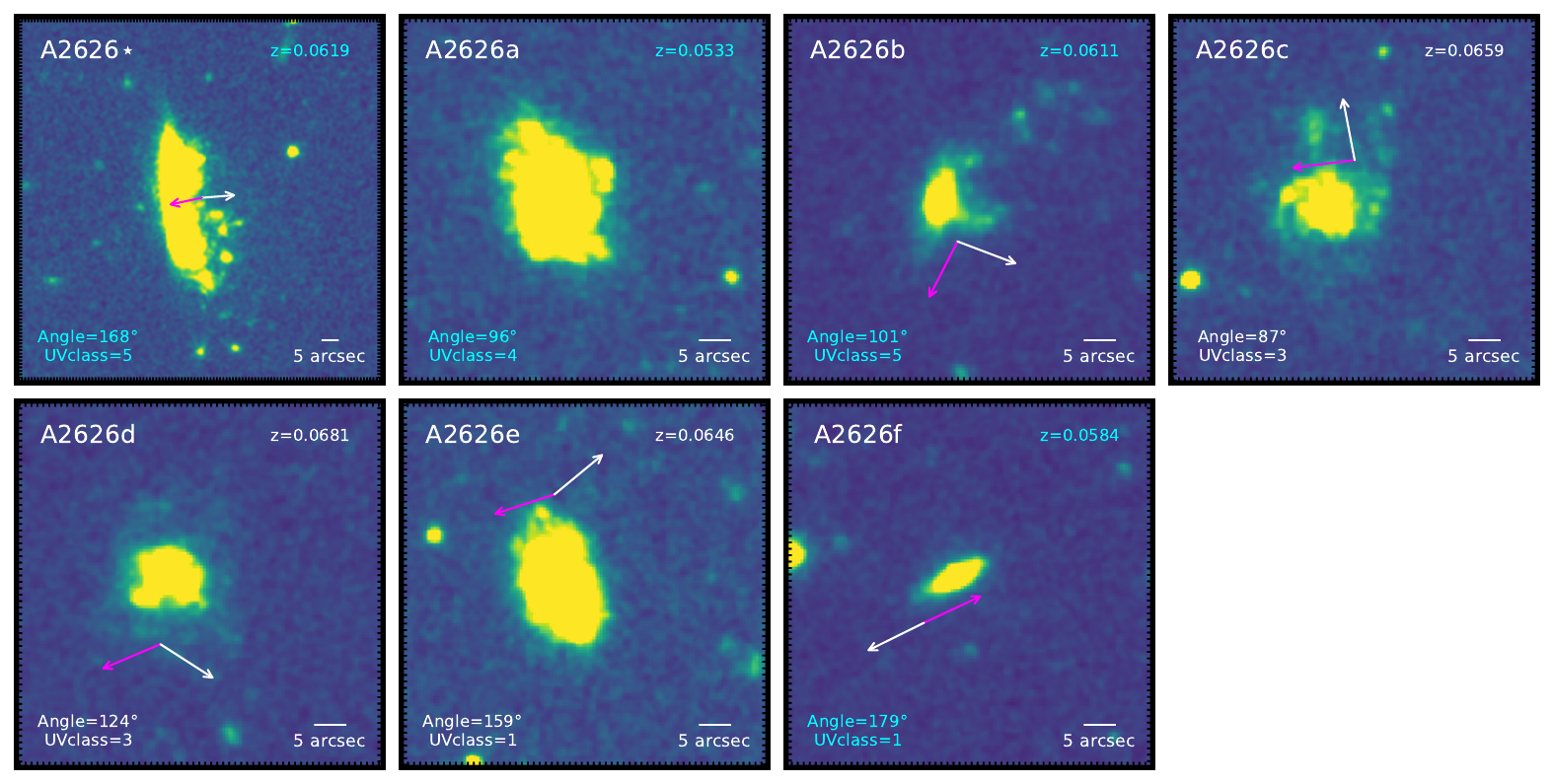}}
\caption{NUV colour scale images of 7 jellyfish galaxies belonging to Abell 2626 galaxy cluster. The 5$\arcsec$ scale bar shown corresponds to 5.32 $kpc$ at cluster rest frame. The details are same as in Fig. \ref{figure:A1668RGB}.}\label{figure:A2626RGB}
\end{figure*}

\begin{figure*}
\centerline{\includegraphics[width=0.53\textwidth]{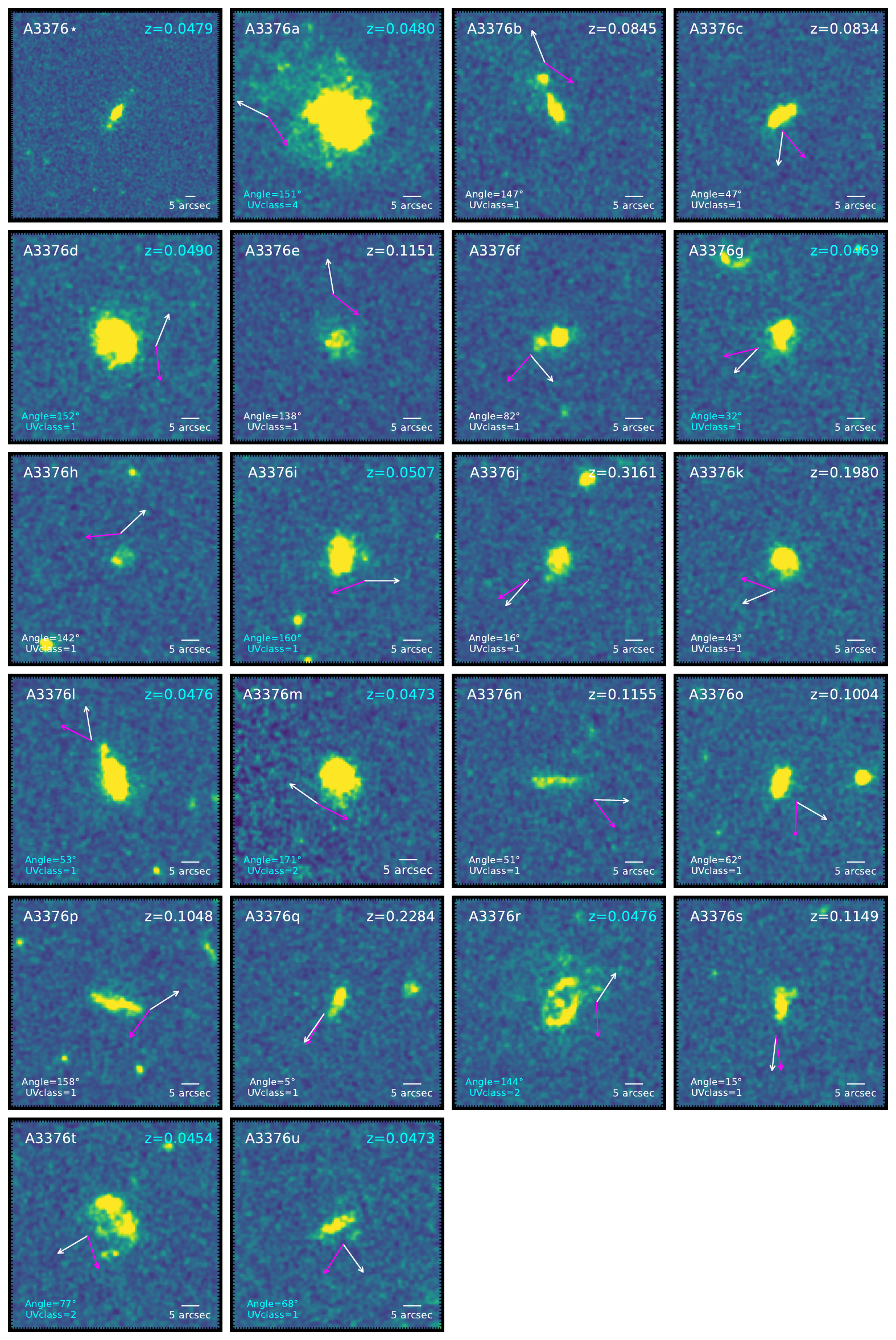}}
\caption{FUV colour scale images of 22 jellyfish galaxies belonging to Abell 3376 galaxy cluster. The 5$\arcsec$ scale bar shown corresponds to 4.52 $kpc$ at cluster rest frame. The details are same as in Fig. \ref{figure:A1668RGB}.}
\label{figure:A3376RGB}
\end{figure*}

\begin{figure*}
\centerline{\includegraphics[width=0.53\textwidth]{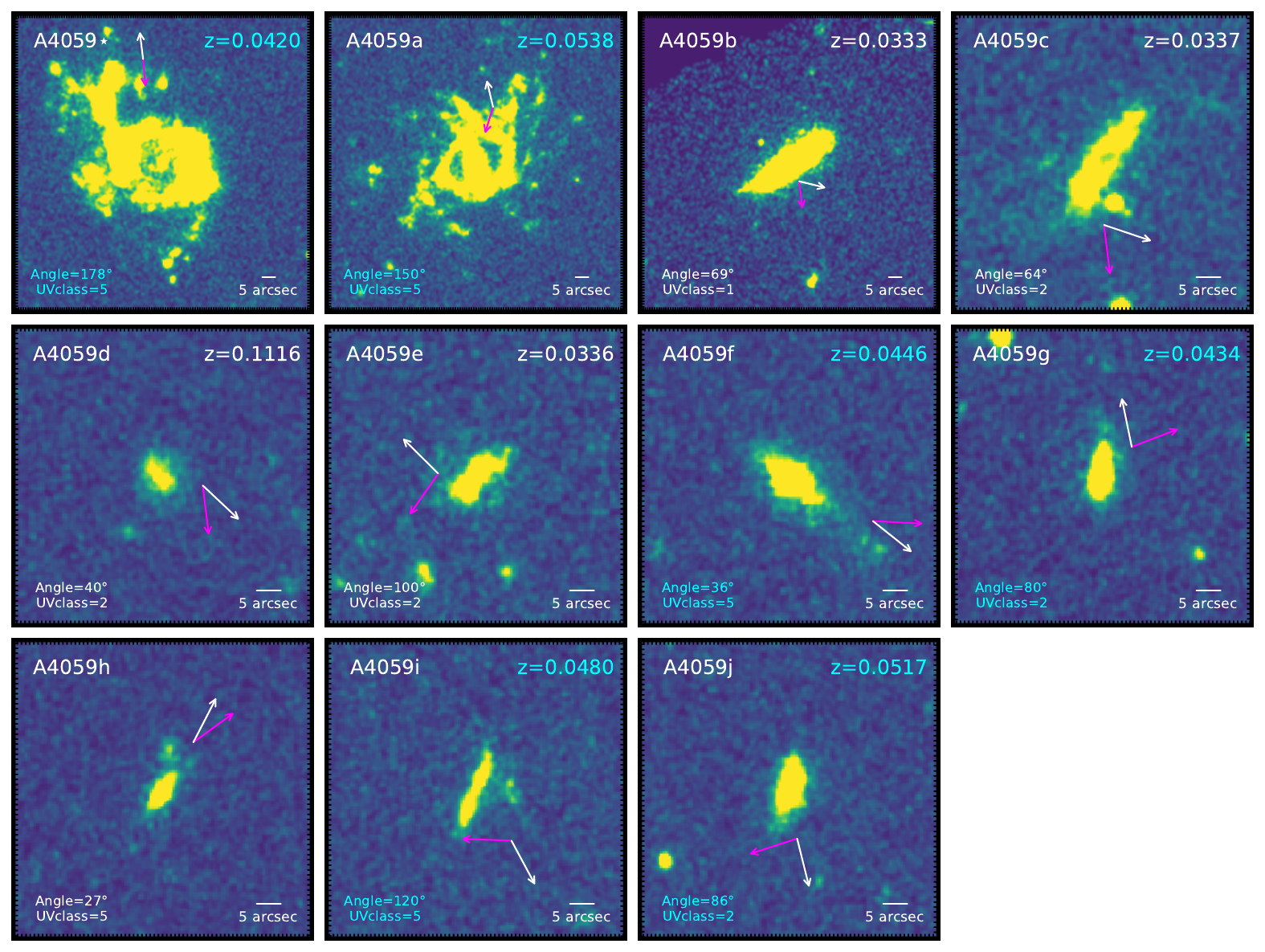}}
\caption{FUV colour scale images of 13 jellyfish galaxies belonging to Abell 4059 galaxy cluster. The 5$\arcsec$ scale bar shown corresponds to 4.70 $kpc$ at cluster rest frame. The details are same as in Fig. \ref{figure:A1668RGB}.}
\label{figure:A4059RGB}
\end{figure*}

\begin{figure*}
\centerline{\includegraphics[width=0.53\textwidth]{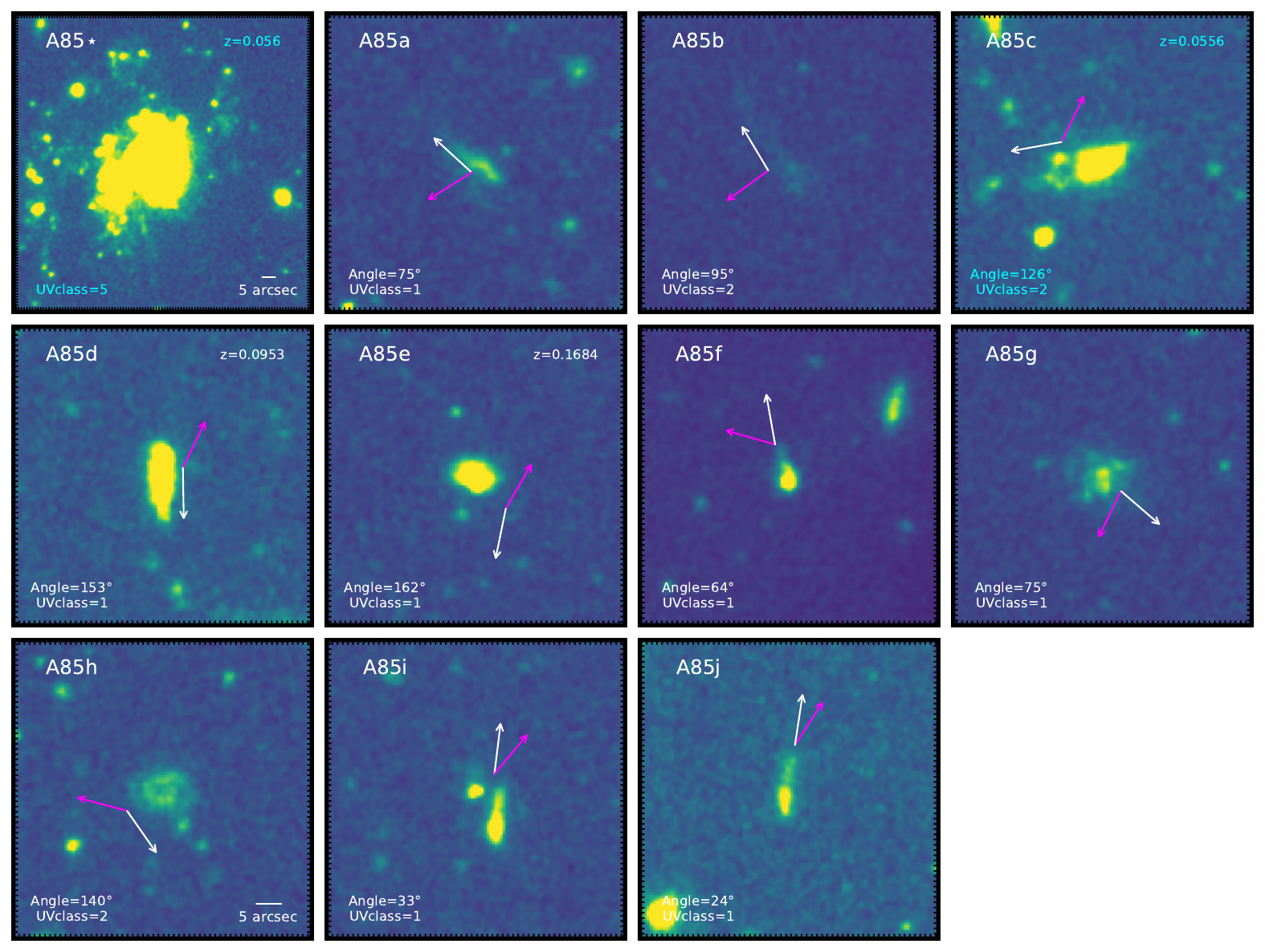}}
\caption{NUV colour scale images of 11 jellyfish galaxies belonging to Abell 85 galaxy cluster. The 5$\arcsec$ scale bar shown corresponds to 5.1 $kpc$ at cluster rest frame. The details are same as in Fig. \ref{figure:A1668RGB}.}\label{figure:A85RGB}
\end{figure*}

\end{appendix}

\end{document}